# An ablating super-Earth in an eccentric binary from the Dispersed Matter Planet Project

John R. Barnes[1], Carole A. Haswell[1], Daniel Staab[1,2], Guillem Anglada-Escudé[3], Luca Fossati[1,4], James P. J. Doherty[1], Joseph Cooper[1], James S. Jenkins[5], Matías R. Díaz[4], Maritza G. Soto[2,4], Pablo A. Peña Rojas[4]

**Earth mass exoplanets are difficult to detect. The Dispersed Matter Planet Project (DMPP) identifies stars which are likely to host the most detectable low mass exoplanets. DMPP-3 (HD 42936) shows signs of circumstellar absorption, indicative of mass loss from ablating planets. Here we report the radial velocity (RV) discovery of a highly eccentric 507 d binary companion and a hot super-Earth planet in a 6.67 d orbit around the primary star. DMPP-3A is a solar type star while DMPP-3B is just massive enough to fuse hydrogen. The binary, with semi-major axis 1.22 ± 0.02 AU, is significantly tighter than others known to host planets orbiting only one of the component stars. The configuration of the DMPP-3 planetary system is rare and indicates dynamical interactions, though the evolutionary history is not entirely clear. DMPP-3Ab is possibly the residual core of a giant planet precursor, consistent with the inferred circumstellar gas shroud.**

Only about 5% of exoplanets are known to reside in multiple star systems, but the very nearest exoplanet is in a triple system[1-3]. Furthermore, 89% of planets with measured masses in multiple star systems are more massive than Neptune[2], while Kepler results imply that sub-Earth to super-Earth sized planets are the most common generally[4]. Thus, there are probably many undiscovered low mass planets in multiple star systems. However, Kepler statistics also imply that the planet occurrence rate in binary stars with orbital separation < 47 AU is only a third that of wider binaries or single stars[5]. Planets in binary star systems fall into two categories: circumbinary (P-type) planets and S-type planets, which orbit a single star in binaries. The latter are found[2] in long period binary systems, with 10 AU < $a_{bin}$ < 28,000 AU. This is unsurprising as tidal truncation of the protoplanetary disc and dynamical perturbations that increase the relative speeds of planetesimals are expected in close binaries[6]. The planetary system we present here, DMPP-3, is a unique discovery: it contains a super-Earth, S-type planet orbiting the primary star in an eccentric binary with orbital semi-major axis of only 1.22 ± 0.02 AU. DMPP-3 thus offers a valuable opportunity to empirically examine dynamical interactions which might create this configuration, for example through the Kozai-Lidov mechanism. Our observations are part of the Dispersed Matter Planet Project (DMPP overview article, *this NA issue*), and were motivated by the anomalously depressed chromospheric emission of the primary star. In cool stars devoid of active regions, Ca II H&K line core emission is powered by non-radiative heating in the chromosphere[7]. This creates a basal level of the activity indicator $\log(R'_{HK})$ = -5.1. A sub-basal value indicates circumstellar absorption, attributable to gas ablated from a close-orbiting planet, such as the super-Earth DMPP-3Ab.

DMPP-3A is HD 42936: a slowly rotating, metal rich, bright (V=9.09) K0V star at a distance of 48.9 ± 0.6 pc (Table 1)[8-12]. It has $\log(R'_{HK})$ = -5.14, and is just below the basal level of stellar activity expected for old, inactive, main sequence stars. We thus selected DMPP-3 as a DMPP target

---

[1] School of Physical Sciences, The Open University, Walton Hall, MK7 6AA Milton Keynes, United Kingdom
[2] AVS, Rutherford Appleton Laboratory, Harwell, Oxford, OX11 0QX United Kingdom
[3] School of Physics and Astronomy, Queen Mary University of London, 327 Mile End Rd, E1 4NS London, United Kingdom
[4] Space Research Institute, Austrian Academy of Sciences, Schmiedlstrasse 6, A-8042 Graz, Austria
[5] Departamento de Astronomía, Universidad de Chile, Camino del Observatorio 1515, Las Condes, Santiago. Chile

DMPP attributes sub-basal activity to absorption by circumstellar gas ablated from undiscovered, close-in, probably low mass planets (see DMPP overview article). We therefore augmented the few existing, sparsely sampled radial velocity (RV) measurements of DMPP-3 (CHEPS project[13]) with high cadence RV observations using HARPS (DMPP project) and two CORALIE observations between 2016 and 2018 (see Methods).

## An 80 $M_J$ companion in an eccentric orbit and a short period planet signal

The RVs span ~1650 m s$^{-1}$ (Fig. 1a) which suggests the presence of a low-mass stellar binary companion orbiting DMPP-3. Various period finding techniques[14-17] reveal periods of 500 d < $P_{orb}$ < 510 d (Fig. 2). The two CORALIE points confirm and constrain the portion of the RV curve around periastron where no HARPS data points yet exist, even though the CORALIE RV uncertainties of 9 ms$^{-1}$ are an order of magnitude greater than the HARPS RV uncertainties. The corresponding maximum *a posteriori* fit, residual RVs and phase folded orbit are shown in Fig. 1a&b. DMPP-3B is a very low mass star, with $M_p \sin i = 79.9^{+3.7}_{-3.0}$ $M_J$ in a $P_{orb} = 506.84^{+0.01}_{-0.02}$ d ($a$ = 1.221 ± 0.023 AU), highly eccentric orbit, with $e = 0.594^{+0.005}_{-0.003}$ (Table 2, Signal 1).

We find a significant low-amplitude Signal 2 when fitted simultaneously with Signal 1 / DMPP-3B[18]. Both the generalized Lomb-Scargle (GLS)[14] and log likelihood periodograms in Fig. 2d,e confirm this signal at $P$ ~ 6.67 d with FAP of 4.0 × 10$^{-5}$. We find maximum *a posteriori* $P_{orb} = 6.6732^{+0.0011}_{-0.0003}$ d and $K$ = 0.97 m s$^{-1}$ with $\Delta \log L$ = 21.4 from the favoured solution with eccentricity, $e$ = 0140[19]]. This indicates a super-Earth planet with $M_p \sin i = 2.58^{+0.35}_{-0.58}$ $M_\oplus$ (Table 2, Signal 2). The phase folded signal is shown in Fig. 1c.

In addition to DMPP-3Ab, the periodograms reveal further significant short period peaks with $P$ = 5.85, 10.40, 14.95 d which are only 5.0, 2.6 and 3.5 times less likely. Further observations will enable assessment of the significance of these peaks. There is also evidence for a longer period ~800 d signal that is likely connected with activity (Fig 2d&e).

Stellar Activity
Although the activity of DMPP-3 indicated by the Ca II H&K line cores is below the basal level, we attribute this to absorption by a diffuse circumstellar cloud of gas ablated from the close-in planet(s) (DMPP overview article, this issue). Consequently, we investigated the possibility that Signal 2 could arise from stellar activity by searching for periodicity in S-index, line full width at half maximum (FWHM) and line bisector span (BIS). The log likelihood periodograms (Fig 3, red solid line) for S-index and FWHM both show a peak close to the period of Signal 1. In the case of S-index, there is a local peak at $P_{orb}$ = 976.9 d, while a local peak in FWHM is present at $P_{orb}$ = 792.6 d. When a long term linear trend is included in the period search (blue dashed line), these peaks become less pronounced. Many of the peaks in both the S-index and FWHM periodograms closely match those seen in the window function (see Supplementary Information for further information on the likely rotation period of DMPP-3A). The BIS shows no clear periodicity, consistent with the low chromospheric activity (i.e. no strong starspot signatures), with FAPs > 10% for all peaks.

The correlation between each of the activity indices and the RVs, after removal of the DMPP-3 B fit is shown in Fig. 4. We subtracted the refined DMPP-3B orbital modulation, Signal 1, while leaving the simultaneously fitted Signal 2 in the residual RVs. These residuals include the full range of the modulations fitted by Signal 2 thus allowing us to examine their corrections with the activity indicators. We find a very weak anti-correlation between S-index and RV residual, with Pearson's $r$ = -0.04. FWHM and BIS show also show weak and very weak positive correlations with RV of $r$ = 0.30 and 0.07. The corresponding F-test p-statistics yield $p$ = 0.68, 3.7 × 10$^{-3}$ and 0.52, suggesting that only the weak FWHM correlation is significant.

Normally we would expect an anti-correlation between BIS and RV if substantial starspots are present on DMPP-3A. Although the lack of correlation is not conclusive, the *r* value suggests starspots of significant size or persistence are absent. Late-type stars exhibit convective blueshift, which is caused by a stronger contribution to emission from the upwelling hotter plasma within a convection cell compared to the cooler down-flowing plasma at the cell edges. Activity and the contribution of plage changes the convective blueshift. For the Sun, the change in line-shape arising from this blueshift modulation induces RV variability of 8 - 10 m s$^{-1}$ over a solar cycle[20]. For later spectral types, the convective blueshift is around 2.5 times smaller, while for stars with log$R'_{HK}$ < -5.1, $\Delta$RV < 5 m s$^{-1}$ is expected[21]. After accounting for Signal 1 and Signal 2, the 1.23 m s$^{-1}$ r.m.s. is commensurate with our estimated age, suggesting DMPP-3 is indeed a low-activity star, with additional circumstellar absorption.

A strong correlation between the Ca II H&K S-index and RV is found in solar data[20] while an apparent lack of correlation between S-index and RV residuals for DMPP-3A is seen. The correlation at 792.6 d between FWHM and RV residuals results from the higher FWHM values seen in the 2018 observations and is consistent with a number of possible contributing effects including the stellar activity cycle, changes in shrouding material, and the presence of long period planets. Therefore, without a longer temporal baseline, we cannot rule out that active regions may be present on DMPP-3A and are responsible for the ~800 d signal seen in Fig. 3.

The properties of DMPP-3B: a hydrogen-burning limit object
With a minimum mass of $M \sin i = 79.9^{+3.7}_{-3.0}$ M$_J$ or $0.076^{+0.004}_{-0.003}$ M$_\odot$ (Table 2, Signal 1) DMPP-3B is consistent with an object exactly at or just above the minimum mass required to sustain hydrogen burning: 0.075 M$_\odot$ or 78.6 M$_J$[22,23]. DMPP-3B is likely to be an L dwarf star with sustained hydrogen burning.

Stars with masses below 0.3 M$_\odot$, corresponding to spectral types M3V-M4V, are fully convective and those with spectral types of M7V or later are likely to exhibit dust condensation, non-equilibrium chemistry, time-varying clouds and molecular opacities. All these characteristics are difficult to model. Furthermore, the very lowest mass stars like DMPP-3B have degeneracies in their observed characteristics, with lower mass, young objects having similar effective temperature and luminosity to higher mass, older objects. Observations of the very lowest mass stars are hampered by their extremely low luminosities: $L \sim 10^{-4}$ L$_\odot$ for mass ~0.08 M$_\odot$. The sample of known objects in this mass regime was very limited[24] but significant progress has been made in recent years. *Gaia* DR2 and SDSS data defined a sample of over 74,000 M and L dwarfs and thus a well-populated empirical colour magnitude diagram of the lower main sequence and its extension to the brown dwarf regime[25]. Of course, the properties of individual objects within this sample are poorly known, though statistical techniques can be used to derive most probable values of age, mass and other properties from the observed photometry. In parallel, a number of benchmark objects have been discovered and intensively observed. For example "benchmark" L dwarfs have been discovered in wide binaries[26-28], moderate period binaries[29], the Coma Berenices cluster where members down to masses of ~0.035 M$_\odot$ have been validated and characterised[30], and finally young, nearby field L dwarfs including 2MASSJ15525906+2948485 and 2MASSJ02411151-0326587 have been characterised[31]. These last two objects may be isolated very low mass stars similar to DMPP-3B. L dwarfs in binaries will be crucial to empirically calibrate the models, exploiting the dynamical mass and metallicity determinations made possible by their brighter binary companions. Many recent discoveries are by-products of transit search programmes and are consequently in short period orbits of < 10 d where they experience significant irradiation from the primary star[32,33]. Such objects are not particularly suitable for calibrating models of isolated analogues. With properly calibrated models, the fundamental parameters of isolated objects can be inferred more reliably, or at least with better understood uncertainties, from the observations. DMPP-3B is a vitally interesting and important object for this reason, and because its orbital eccentricity means we can potentially probe how the atmospheric properties respond to changing proximity of the primary star. DMPP-3AB is not a particularly tight binary system: even with $e$ = 0.597, tidal dissipation between the stars is negligible, with inspiral time $\tau_B \sim 1.2 \times 10^{17}$ yrs[34].

We estimate the age of DMPP-3 at ≥ 6.2 Gyr (Table 1). For ages 2 - 10 Gyr, there is little change in luminosity expected for a 0.08 $M_\odot$ object[35]; we expect the V-band contrast ratio between DMPP-3A and DMPP-3B to be $(2 - 3) \times 10^6$. At a distance of 48.9 pc[11], the semi-major axis of the orbit is equivalent to 0.025″ and is thus not resolved by standard photometric surveys. DMPP-3B might, however, be resolvable by state-of-the-art imaging systems and thus offers the prospect of measuring the emission from a binary companion at the very bottom of the main sequence[36]. High resolution spectroscopic techniques that make use of the many photospheric lines should be able to detect DMPP-3B in the photometric K band[37] where it is expected to be only ~800 - 1000 times fainter than DMPP-3A. The estimated masses of DMPP-3A and DMPP-3B imply a DMPP-3B velocity amplitude of $K \sim 30 \pm 2$ km s$^{-1}$ (for orbital inclination, $i$ = 90º), ensuring the spectroscopic signatures of each component are well resolved. Astrometric observations by *Gaia* should enable us to determine the true mass and thus the orbital inclination[38].

The super-Earth planet: probability of transits, eclipses and phase curve variations

The relatively low amplitude of the significant 6.6732 d Keplerian Signal 2 is reflected in the 18%, uncertainty of the derived minimum mass. For this period, for random orientations the transit probability is 6.4%[39]; however, angular momentum considerations suggest the ablated planetary material is likely to remain concentrated in the orbital plane. Consequently, the transit probability for bodies in the DMPP-3 system is higher than for a randomly oriented system.

For randomly oriented orbits, the probability that the DMPP-3AB system is an eclipsing binary is small. The distance between the two stars at inferior conjunction of DMPP-3B is 1.09 AU, implying a probability of transit of only 0.4% if randomly oriented. Nonetheless the possibility of an eclipsing binary containing a star at the hydrogen-burning limit is exciting, and worth exploring with high quality photometry.

The proximity of the DMPP-3 system, and the apparent lack of star spots on DMPP-3A makes DMPP-3Ab an excellent prospect for detection of phase-dependent reflected light. This effect has less demanding alignment requirements than transits, and the existence of absorbing material in the line of sight implies the system is likely to be more-or-less edge-on. For this reason, and to search for transits and eclipses DMPP-3 is an excellent prospective target for high quality space-based photometry, but DMPP-3 is in a region of sky inaccessible to the European Space Agency's CHEOPS mission (European Space Agency's CHEOPS mission webpage at http://sci.esa.int/cheops).

Stability of DMPP-3Ab and implications of orbital simulations for empirical RVs

With $P_{orb}$ = 505.6 d and $e$ = 0.63, the period and eccentricity of HD 191760's companion are almost identical to DMPP-3B's, though $M_p \sin i$ = 38.2 $M_J$[13] places HD 191760B firmly in the brown dwarf regime. Although no planet candidates are identified, HD 191760B constrains possible orbital configurations and stability for planets interior to its orbit: orbits at > 0.17 - 0.18 AU are not expected to be stable. Similarly, a study investigating the orbital stability of planets in binary systems with different mass ratios[40] predicts that interior planets (i.e., S-type planets orbiting one binary component) are stable for $10^4$ binary orbits with planetary semi-major axes $a_p$ < 0.16 AU. To test this for DMPP-3, we carried out orbital integrations for our solution (Table 2 and methods for further details). The system remained stable for $10^6$ yrs or equivalently >720,600 orbits of the DMPP-3AB binary; a characteristic quasi-periodic modulation timescale of 800 yrs is found to repeat throughout the simulation. The eccentricity of DMPP-3Ab is modified by the orbit of DMPP-3B. Starting the simulation with $e_{p,start}$ = 0.14 results in modulations of the eccentricity with mean values per orbit of 0.00 < $e_p$ < 0.18. The value of $e_p$ varies by up to 0.13 on the orbital timescale of DMPP-3B, although typically $e_p$ varies with an r.m.s. of 0.02.

In all orbital simulations the DMPP-3B orbit pumps the eccentricity of the inner planet to higher values; even a starting eccentricity of $e_{p,start} = 0$ results in similar 800 yr quasi periodic behaviour as described above, with $0.00 < e_p < 0.16$. Although the eccentricity of the inner planet may be negligible at times, the modulation to higher values means that at any given instant, the eccentricity is likely to be greater than zero. With $e_{p,start} = 0$ and 0.14, bimodal distributions of $e_p$ are found in each simulation with respective most probable peaks at $e_p = 0.038$ and 0.063 while the second peaks are respectively found at $e_p = 0.121$ and 0.137. The fit to our observations with $e_p > 0$ is thus more likely; forcing $e = 0$ solutions would only arise if we happened to observe at a rare and special epoch since $e < 0.01$ for only 2.8% of the time during the simulations with $e_{p,start} = 0.14$.

The variability in $a_p$ of 0.9% is small but the implied 1.4% changes in orbital period due to DMPP-3B – planet interactions could potentially cause significant side-peaks in periodograms when observations are taken at multiple epochs. This will decrease the inferred significance of detected periodicities. The orbital stability simulations reveal that the 0.01% period uncertainties in Table 2 are underestimated. As expected, the simulations show perturbations in $a$, $P_{orb}$ and $e$ for DMPP-3B are two orders of magnitude smaller than for DMPP-3Ab.

Formation and Evolution of the DMPP System
DMPP-3 is a dynamic system in a potentially important stage of its evolution, comprising a K0V star, an object at the hydrogen-burning limit in an eccentric 507 d binary orbit, and a close-orbiting circumprimary (S-type) super-Earth planet. It is the only super-Earth S-type binary planet detected via the radial velocity method orbiting an FGK star. Including DMPP-3Ab, of the 75 known S-type systems with estimated masses[41-44], only 5 are found with $a_{bin} < 10$ AU. The next tightest system, ν Oct b, with $a_{bin} \sim 2.6$ AU harbours a $\sim 2.1$ M$_{Jup}$ planet in a $\sim 1.3$ AU retrograde orbit[44]. DMPP-3 is thus by a factor of 2 the tightest known binary with an S-type planet. As such it presents a challenge to models of planet formation[6]. Conventional models predict that any circumprimary protoplanetary disc would be truncated by the tidal effect of DMPP-3B, which would both reduce the mass available for formation of planets and severely limit the lifetime of the disc. Furthermore, planetesimals forming in the protoplanetary disc are liable to be excited into high eccentricity orbits through interactions with DMPP-3B and thus planetesimal-planetesimal encounters will usually occur with high relative velocity, inhibiting aggregation. These difficulties almost certainly underlie the dearth of known S-type planets in short period ($P_{orb} < 2000$ d) binaries[41].

An alternative formation model is capture of circumbinary (P-type) planets into S-type orbits following scattering. Simulations suggest only $\sim 1\%$ of scattered P-type planets would be so captured in DMPP-3 so this also seems intrinsically unlikely[6].

The most promising mechanism for the current proximity of DMPP-3Ab to its host star is the eccentric Kozai-Lidov (KL) effect[45], which can excite high degrees of eccentricity for the inner orbit in a triple system. In this case, the orbit of DMPP-3Ab about DMPP-3A is perturbed by the eccentric orbit of DMPP-3B. The planet's eccentricity is modulated on the KL timescale. The planet may merge with the star during times of very high eccentricity; however, if the tidal timescale is similar to or shorter than the KL timescale[46], the planet orbit will circularise and thus then shrink. The planet must still form or arrive in an orbit interior to DMPP-3B, so tension remains with planetary system formation models. Recent work suggests isothermal fragmentation of supersonically infalling protostellar clouds could provide an abundance of planet mass fragments[47]. Forming planets from these fragments may be problematic, however, once the luminosity of the central protostar negates the isothermal assumption. In addition, the formation of very low mass stars like DMPP-3B is poorly understood and a subject of ongoing debate[28].

# Discussion and Concluding Remarks

DMPP-3 was selected as a DMPP target because of its anomalously low log$R'_{HK}$, suggesting the presence of enshrouding gas from a mass losing planet. This target selection enables exciting efficient detections of close-orbiting, low-mass planets with HARPS. DMPP planets may reveal a rare, short-lived phase of evolution during, or immediately after, significant planetary mass-loss. In DMPP-3AB we have an opportunity to directly observe the eccentric Kozai-Lidov mechanism. The behaviour of such systems is *'exciting, rich and chaotic in nature'* [45]. Most currently known low-mass planets are located below the lower boundary of the Neptune desert[48] and may have already undergone significant mass-loss: i.e. they may be chthonian planets, the remnant cores of giant planets. For example, the warm Jupiter system Kepler-693 is potentially a similar system to DMPP-3, but in an earlier evolutionary phase[42]. DMPP-3Ab is a nearby super-Earth below the lower mass boundary of the Neptune desert. While the size of DMPP-3Ab is unknown, its mass suggests it lies below the radius valley[49]. If it is the source of the shrouding material, it may be an unusually low gravity planet and it may reveal the process creating the desert. The equilibrium temperature of DMPP-3Ab is $T_{eq}$ = 854 K (assuming a Bond albedo of 0.25), too low for surface liquid magma, but potentially sufficient to generate a stellar-wind sputtered exosphere[50] (see also DMPP-1 article in this issue of NA, which discusses this issue). Furthermore, tidal stresses due to the pumped eccentricity of its compact orbit will provide additional heating of DMPP-3Ab. DMPP-3A may also host one or more lower mass planets, interior to the DMPP-3Ab orbit. Further RV observations can probe this. In DMPP-3, the mass loss phenomenon seems likely to be related to KL migration.

With 101 observations we have identified the candidate planet DMPP-3Ab. Our simulations of orbital stability indicate that the potential benchmark low-mass DMPP-3B induces variable and cyclic values of eccentricity in the planet's orbit. Although the binary orbit is established, precision RV observations around periastron would significantly improve our knowledge of the orbital elements. This would also help to further constrain the periodicity of the ~800 d putative activity signal. Intensive monitoring spanning more than one orbit of DMPP-3Ab might enable any changes in eccentricity to be monitored if observed at times of apastron and periastron of DMPP-3AB.

**Table 1**. HD 42936 (DMPP-3A) stellar parameters and references (for SIMBAD see http://simbad.u-strasbg.fr).

| Parameter | Value | Ref |
|---|---|---|
| Spectral Type | K0V | 8 |
| Parallax [mas] | 20.45 ± 0.26 | 9 |
| Distance [pc] | 48.9 ± 0.6 | 9 |
| $V$ [apparent mag.] | 9.09 | SIMBAD |
| $B-V$ [apparent mag.] | 0.91 | SIMBAD |
| log($R'_{HK}$) | -5.14 ± 0.05 | 11 |
| $T_{eff}$ [K] | 5138 ± 99 | 12 |
| [Fe/H] | 0.18 ± 0.09 | 10 |
| log $g$ [cm s$^{-2}$] | 4.30 ± 0.51 | 12 |
| $v \sin i$ [kms$^{-1}$] | 1.97 ± 0.14 | 12 |
| $v_{mac}$ [kms$^{-1}$] | 1.50 ± 0.14 | 12 |
| $R_*$ [$R_\odot$] | 0.91 ± 0.02 | 12 |
| $M_*$ | 0.87 ± 0.05 | 12 |
| Age [Gyr] | 10.9 ± 4.7 | 12 |

**Table 2.** Solution with two simultaneously fitted Keplerians. The Signal 1 maximum *a posteriori* parameters describe the DMPP-3B orbit while the favoured Signal 2 (with eccentricity prior of $0.0 < e < 0.80$) parameters reveal the super-Earth planet DMPP-3Ab. The Generalized Lomb-Scargle false alarm probabilities (FAPs) are given along with the maximum *a posteriori* improvement in log likelihood, $\Delta \log L$, compared with the respective no-companion and Signal 1 models. The reduced $\chi_r^2$ (calculated by including the excess white noise terms, $\sigma_{CHEPS}$, $\sigma_{DMPP}$ and $\sigma_{CORALIE}$) and the fit r.m.s are also tabulated. The 68.3% confidence ranges (in parentheses) are obtained from MCMC sampling. The mean anomaly, $M_0$, and longitude of periastron, $\omega_0$, are given for the reference epoch of $t_0$. The instrumental velocity offsets are $\gamma_{CHEPS}$, $\gamma_{DMPP}$ and $\gamma_{CORALIE}$. Signal 2 values in square parentheses are for a forced $e = 0$ fit. The change in Bayesian Information Criterion value as successive signals are added ($\Delta$BIC) indicates there isn't sufficient evidence to prefer the $e = 0.140$ solution over the $e = 0$ solution.

| Parameter | Signal 1<br>DMPP-3B | Signal 2 [$e = 0$ solution]<br>DMPP-3Ab |
|---|---|---|
| FAP (GLS) | $6.7 \times 10^{-48}$ | $4.0 \times 10^{-5}$ |
| $\Delta \log L$ | 546.6 | 21.4 [21.1] |
| $\chi_r^2$ | 1.20 | 1.25 [1.26] |
| r.m.s. | 1.46 | 1.23 [1.26] |
| $\Delta$BIC | $3.5 \times 10^5$ | 19.8 [19.1] |
| $P$ [d] | 506.844 (506.828 - 506.859) | 6.6732 (6.6729 – 6.6743) [6.6736] |
| $K$ [m s$^{-1}$] | 2627.9 (2595.3 - 2687.3) | 0.97 (0.78-1.09) [0.95] |
| $M_0$ [deg] | 126.16 (126.00 - 126.29) | 184.78 (126.27 - 226.71) [192.7] |
| $e$ | 0.594 (0.591 - 0.599) | 0.140 (0.087-0.231) [0 - *fixed*] |
| $\omega_0$ [deg] | 158.61 (158.42 – 158.92) | 80.57 (58.93 – 153.41) |
| $\gamma_{CHEPS}$ [m s$^{-1}$] | | -826.6 (-837.1 - -820.2) |
| $\gamma_{DMPP}$ [m s$^{-1}$] | | -819.7 (-830.8 - -813.2) |
| $\gamma_{CORALIE}$ [m s$^{-1}$] | | 3570.2 (3526.0 - 3652.9) |
| $\sigma_{CHEPS}$ [m s$^{-1}$] | | 1.4 (1.3 - 2.8) |
| $\sigma_{DMPP}$ [m s$^{-1}$] | | 0.4 (0.4 - 0.6) |
| $\sigma_{CORALIE}$ [m s$^{-1}$] | | 0.0 (0.0 - 0.0) |
| $M_p \sin i$ | 79.9 (76.9 - 83.6) M$_J$ | 2.58 (2.00 - 2.93) M$_\oplus$ |
| $a_p$ [AU] | 1.221 (1.198 - 1.244) | 0.0662 (0.0649 - 0.0675) |
| $N_{obs}$ ($N_{CHEPS+DMPP+CORALIE}$) | | 8 + 91 + 2 |
| Baseline [day / years] | | 3559 / 9.7 |
| $t_0$ [BJD] | | 2454579.56469 |

## Acknowledgements

This work is based on observations collected at the European Organisation for Astronomical Research in the Southern Hemisphere under ESO programmes 081.C-0148(A), 088.C-0662(A) and 091.C-0866(C), 096.C-0876(A) and 098.C-0269(A), 098.C0499(A), 098.C0269(B), 099.C-0798(A) and 0100.C-0836(A). D.S. was supported by an STFC studentship. CAH and JRB were supported by STFC Consolidated Grants ST/L000776/1 and ST/P000584/1; DS was supported by an STFC studentship. GA-E was supported by STFC Consolidated Grant ST/P000592/1; JSJ acknowledges support by FONDECYT grant 1161218 and partial support from CONICYT project Basal AFB-170002. M.R.D. acknowledges the support of CONICYT-PFCHA/Doctorado Nacional-21140646, Chile and project Basal AFB-170002. These results were based on observations awarded by ESO using HARPS. This research has made use of the SIMBAD data base, operated at CDS, Strasbourg, France.

## Author Contributions

JRB contributed to proposals, performed RV analyses and wrote the paper. CAH leads all aspects of the DMPP collaboration, secured the funding, wrote the proposals, and co-wrote the paper. DS performed target selection, contributed to writing of proposals, initial RV analyses and technical details of the paper. GA-E provided software and expertise. LF contributed to the analysis and proposal writing. JSJ, MGS and PAPJ provided expertise on stellar activity, the log $R'_{HK}$ metric, contributed stellar parameter analyses and performed radial velocity solution checks. DS, CAH, JRB, JD, JC and MRD performed observations with HARPS. All authors were given the opportunity to review the results and comment on the manuscript.

## Author Information

Reprints and permissions information is available at www.nature.com/reprints. Correspondence and requests for materials and codes should be addressed to john.barnes@open.ac.uk.

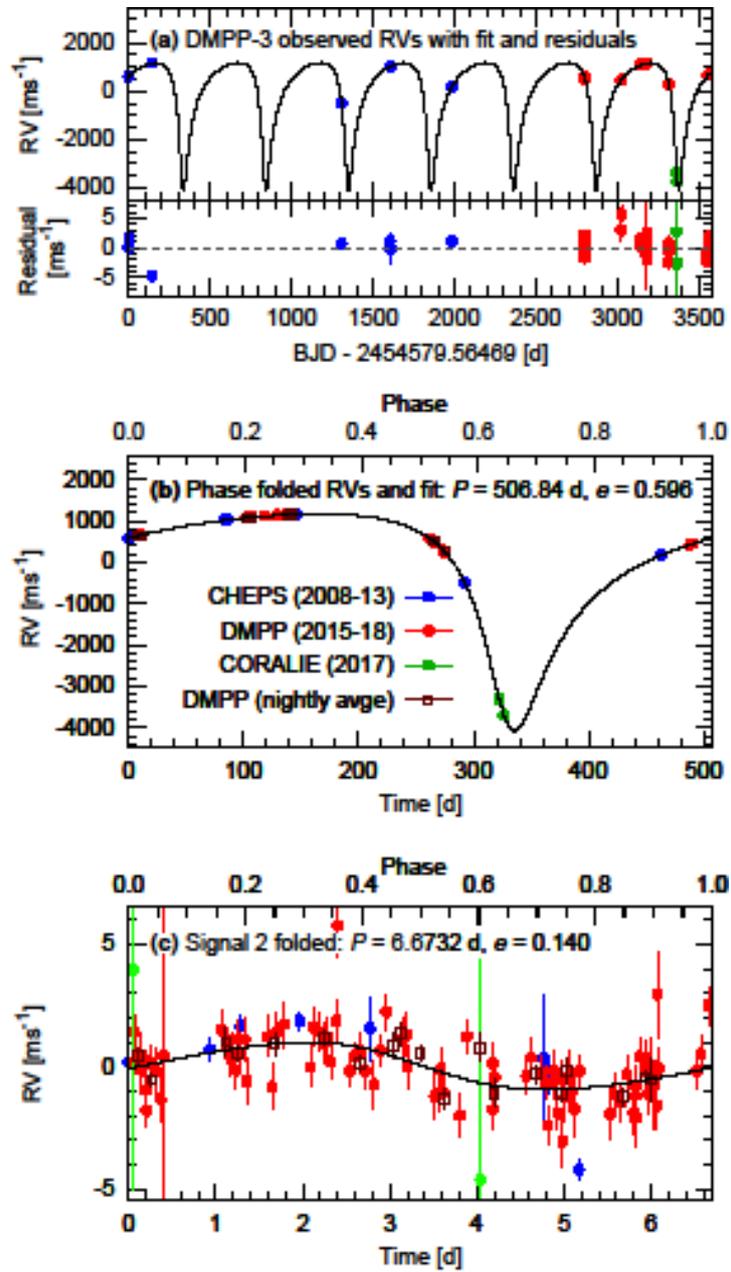

**Figure 1. (a)** DMPP-3 radial velocity observations, maximum *a posteriori* fit[19] (see Table 1) and residuals for the $M_p \sin i \sim 80$ $M_J$ companion DMPP-3B in a high eccentricity orbit **(b)** RVs folded on the DMPP-3B orbit **(c)** Phase fold of Signal 2, indicating a super-Earth DMPP-3Ab. RVs are plotted with 1-σ uncertainties.

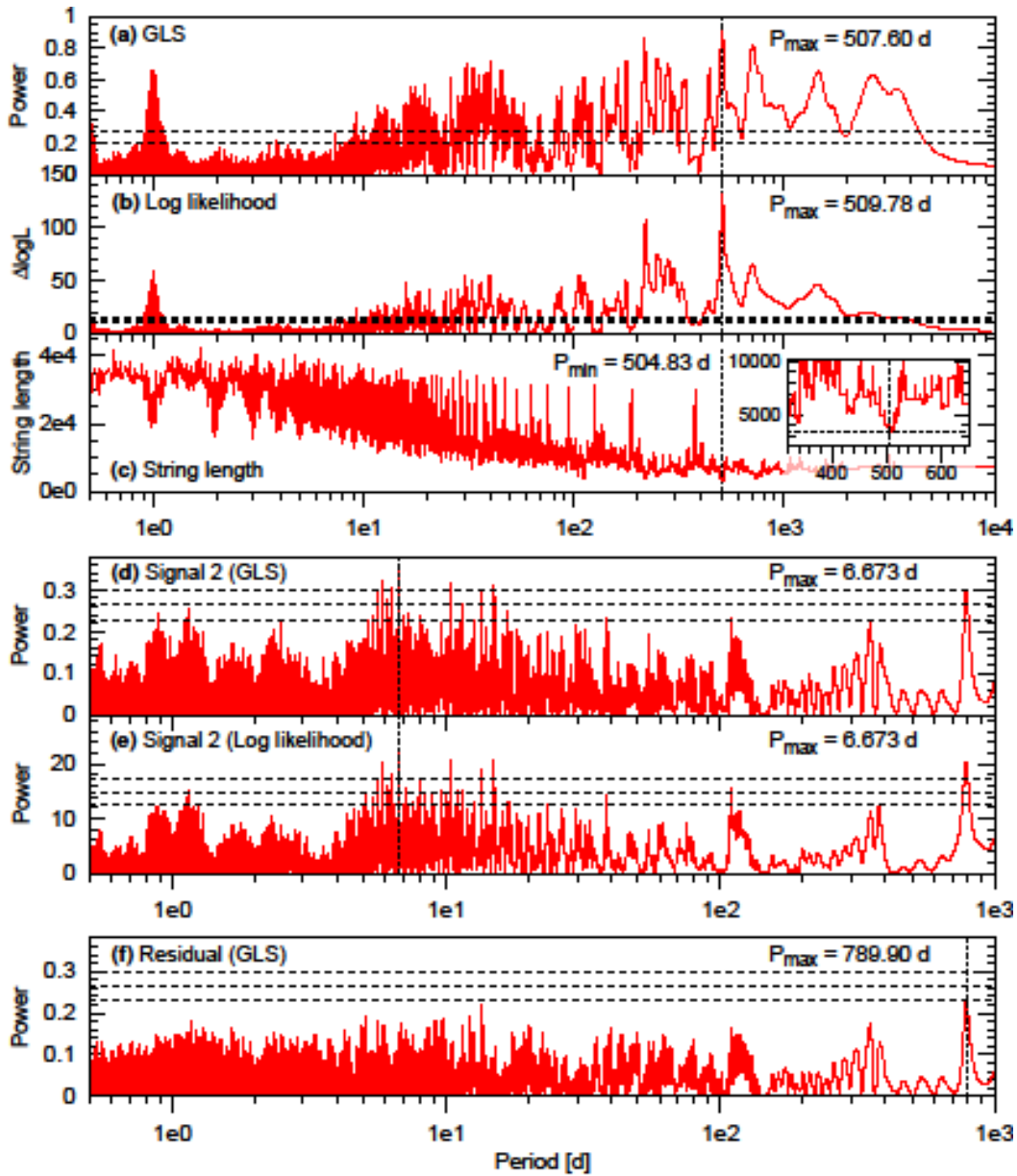

**Figure 2**. **(a)** Generalised Lomb Scargle (GLS) periodogram with maximum at $P_{orb}$ = 507.60 d (FAP = 6.7 × $10^{-48}$). **(b)** Log likelihood periodogram with maximum at $P_{orb}$ = 509.78 d. **(c)** String length as a function of period revealing a minimum orbital period at $P_{min}$ = 504.83 d. **(d)** Signal 2 GLS periodogram and **(e)** Log likelihood periodogram of residual RVs when optimising simultaneously with the binary parameters (Signal 1, Table 2). $P_{orb}$ = 6.673 d and FAP = 4.0 × $10^{-5}$. **(f)** GLS periodogram of residuals following the simultaneous maximum *a posteriori* fit for Signals 1 and 2. No further significant signals remain; the ~800 d signal has 6% FAP. All periodograms show 0.1, 1 and 10% FAPs.

**Figure 3.** Log likelihood periodograms (red lines) of activity indicators: S-index, FWHM and BIS. Blue dashed curves result from period searches with a linear trend included in the model. The window function combined with activity modulations show significant aliasing at periods between 7 d and 500 d. The vertical lines indicate the period of the DMPP-3AB binary and the ~800 d RV and FWHM peaks.

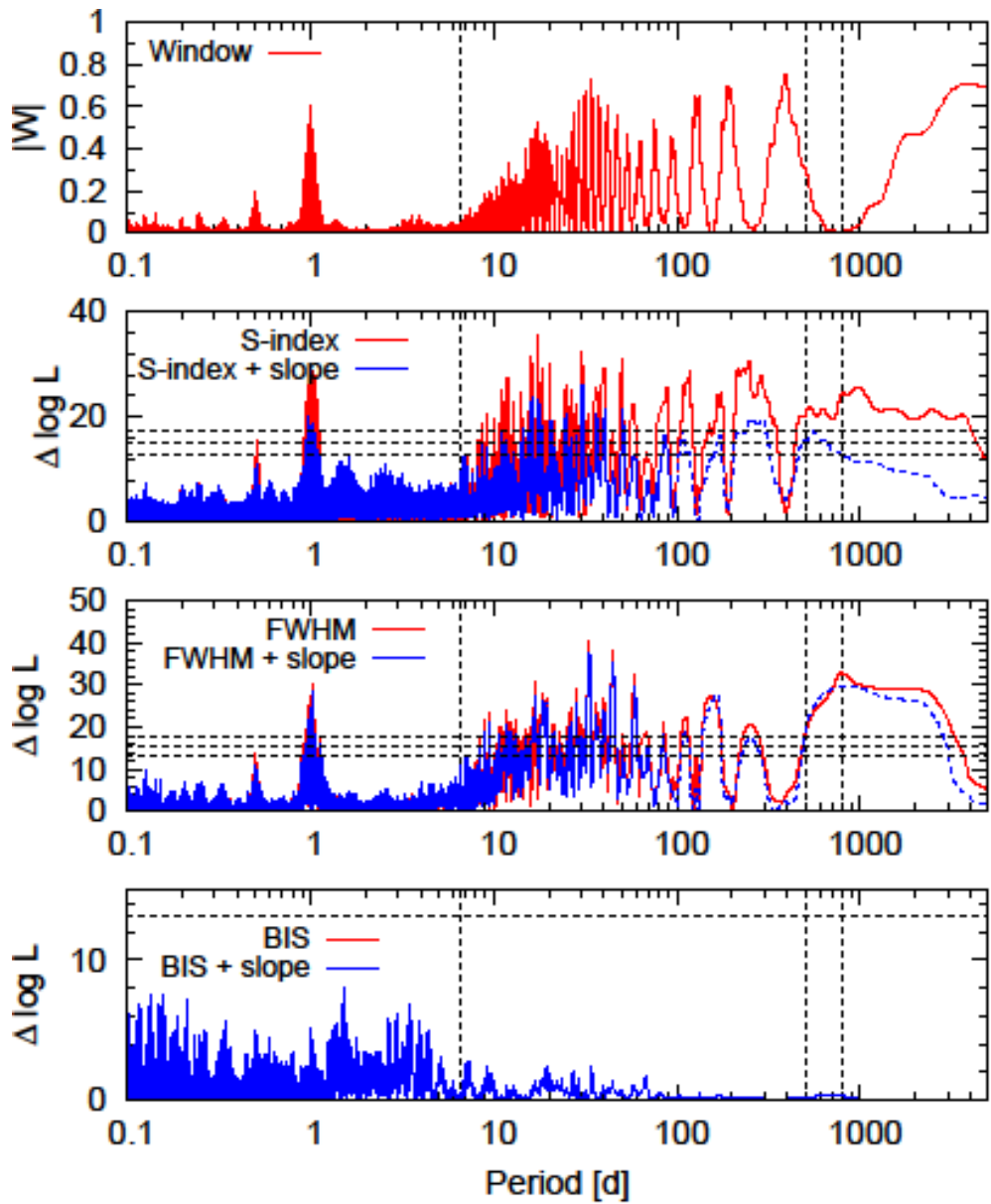

**Figure 4**. Activity indices vs radial velocities (both plotted with 1-σ uncertainties). The S-index, line full width at half maximum (FWHM) and line bisector span (BIS) are plotted against the radial velocities after subtraction of Signal 1 in Table 2.

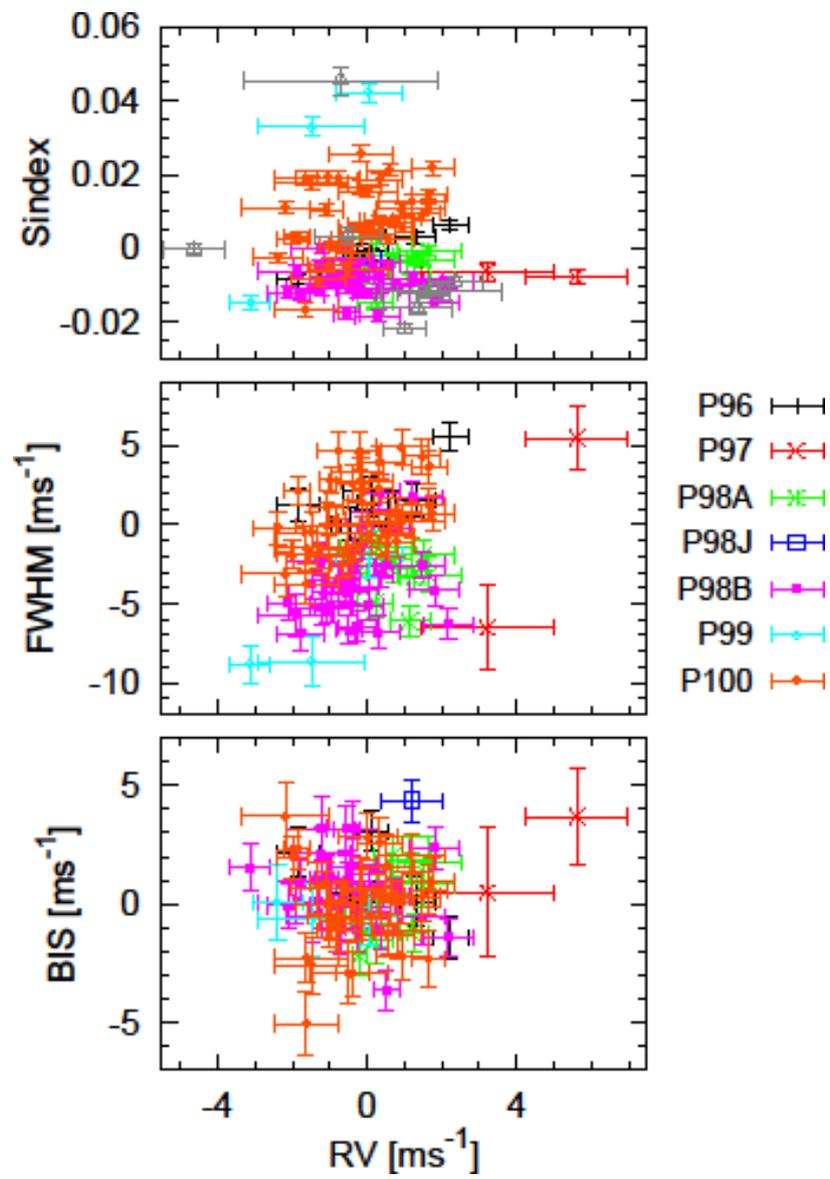

## Methods

### Observations

HARPS RVs of HD 42936 were obtained between 2008 and 2013 in programmes 081.C-0148(A), 088.C-0662(A) and 091.C-0866(C) as part of the Calan-Hertfordshire Extrasolar Planet Search (CHEPS), which targeted high metallicity stars. We refer to these 8 observations as the CHEPS observations. We subsequently and independently identified HD 42936 as a low activity target and have observed it intensively during five further observing campaigns (096.C-0499(A), 097.C-0390(B), 098.C-0269(B), 099.C-0798(A) and 0100.C-0836(A); hereafter P96, P97, P98, P99 and P100) between 2015 and 2018. We collectively refer to these observations as the Dispersed Matter Planet Project (DMPP) data set. Details of the DMPP observing strategy are given in the Methods section of the overview article in this issue of NA. The 098.C-0269(B) / P98 run comprised two sub-runs in 2016 Nov/Dec and 2017 Jan, which we refer to as P98A and P98B. In addition, a time swap agreement enabled two observations of HD 42936 to be made in late 2016 Dec which we refer to as P98J. We also made two further observations with CORALIE at the Swiss 1.2-metre Leonhard Euler Telescope in 2017 July 8 and 12.

### Data Reduction

The RVs were obtained using HARPS-TERRA[51] which uses a template spectrum, derived from the co-addition of spectra. HARPS-TERRA relies on least-squares minimisation of the difference between a high SNR template and individual stellar spectra of a time series. Extracted, wavelength calibrated spectra from the DRS output are the starting point for the template matching approach.

### DMPP-3 / HD 42936 system parameters

Our code, `SPECIES`[12], uses high resolution spectra to determine several of the stellar parameters of DMPP-3 (further details of the are given in the DMPP-1 article of this NA issue).

### Period analysis tools

The RVs measured by CHEPS and DMPP with HARPS span ~1650 m s$^{-1}$ (Fig. 1a) which suggests the presence of a low-mass stellar binary companion orbiting DMPP-3. We employed the generalized Lomb-Scargle (GLS)[14] analysis, which improves the standard Lomb-Scargle period method[52,53] by including data weights and accounting for a data offset (the Python implementation can be found at https://pyastronomy.readthedocs.io/en/latest/pyTimingDoc/pyPeriodDoc/gls.html). A maximum is found in the Generalised Lomb-Scargle (GLS) periodogram[14] at $P_{orb}$ = 507.60 d with false alarm probability (FAP) of 6.7 × 10$^{-48}$). We also used the recursive period finding algorithm, `RECPER`, which maximises the global likelihood each time an additional planet signal is added to the model[15,3]. The maximum in the log likelihood periodogram[15] is at $P_{orb}$ = 509.78 d, with $\Delta \log L$ = 132.7 (FAP = 6.5 × 10$^{-52}$) (Fig. 2a,b). String length minimisation, which makes no assumptions about the RV curve morphology[16,17], yields $P_{orb}$ = 504.83 d (Fig. 2c). This method is particularly useful for orbits that exhibit large degrees of eccentricity. The method involves phasing the observations for a range of trial periods and calculating the sum of the distances between all neighbouring phased points.

A full Bayesian analysis using the EMPEROR code (PAPR and JSJ; in preparation 2019) also reveals a period of a little over 500 d, strengthening the case for this solution. Inspection of the phase-folded RVs with 500 d < $P_{orb}$ < 510 d reveals an eccentric orbit that is poorly sampled at some phases. The global maximum *a posteriori* orbital parameters[19] are listed in Table 1 and Supplementary Table 1 for the binary orbit. The corresponding maximum *a posteriori* fit, residual RVs and phase folded orbit are shown in Fig. 1a&b.

Visual identification and inspection of Keplerian signals was made using Version 2 of the `Systemic Console`[18]. After constraining the binary orbit of DMPP-3AB, we then used `Systemic` to identify further significant signals via Lomb-Scargle periodogram analysis. Additional signals were added to the model and the parameters simultaneously optimised for all signals. After identifying significant signals using `Systemic`, and verifying them with GLS and RECPER periodogram searches, subsequent maximum *a posteriori* parameter determination and parameter uncertainty estimates were obtained using the Radial Velocity Fitting Toolkit, `Radvel`[19]. This is the most robust method of signal identification and parameter optimisation owing to the lack of phase coverage and the large eccentricity of DMPP-3B.

For the activity periodogram analysis, we again used the maximum likelihood routine, RECPER. The window function in Fig. 3 is calculated in `RECPER` via a discrete Fourier Transform[54].

Independent Data Sets and HARPS fibre upgrade offset

The CHEPS observations were all made prior to the HARPS fibre upgrade[55] in 2015, while the DMPP observations were made after the upgrade. We derived RVs for *all* the HARPS observations using *only* the P98B template and then treated the resulting CHEPS and DMPP (and CORALIE) data values as independent data sets. This enables the post-fibre minus pre-fibre systematic velocity offset to be determined. We find an RV offset of $\gamma_{DMPP} - \gamma_{CHEPS} = 6.9$ ms$^{-1}$.

Signal 2 – a planet with eccentricity?

Given the eccentricity of DMPP-3B, it is not clear that Signal 2 should be purely sinusoidal. Therefore, we investigated fits of eccentric signals[19], allowing the orbital elements of DMPP-3B to be refined in the simultaneous fit. For Signal 2, we set an initial guess of $e = 0.1$ during the maximum *a posteriori* fitting and MCMC chains with a uniform prior restricted to the range $0.0 < e < 0.8$. Although eccentricity is favoured, the fit significance and other orbital parameters for $e = 0$ are almost identical (Table 2, $e=0$ values in square parentheses). The corresponding small difference of 0.7 in Bayesian Information Criterion ($\Delta$ BIC) values for the $e = 0$ and $e = 0.140$ cases indicates that neither solution is preferred.

Orbital stability analysis

Stability analyses were performed using the N-body orbital integrator, `REBOUND`[56]. Since *a priori* close encounters might arise due to mutual perturbations of the planets, we used the `IAS15` integrator[57], which uses automatic integration steps.

## Data availability Statement

The data that support the plots within this paper and other findings of this study are available from the corresponding author upon reasonable request.

## Methods section references

[57] Rein, H. and Spiegel, D. S. "IAS15: a fast, adaptive, high-order integrator for gravitational dynamics, accurate to machine precision over a billion orbits" *Mon. Not. R. Astron. Soc.* **446**, 1424-1437 (2015)

## Supplementary Information

Initial single Keplerian solution for DMPP-3B only

The radial velocity (RV) residuals after subtraction of the solution with a single signal (DMPP-3B) are listed in Supplementary Table 1.

The CORALIE observations

Taken 4 days apart, the RV of the second CORALIE observation is 387.7 m s$^{-1}$ less than the first. The solution without CORALIE points results in consistent, but marginally lower eccentricity, $e = 0.593^{+0.007}_{-0.004}$, and semi-amplitude, $K = 2610.5^{+75.1}_{-46.8}$ m s$^{-1}$; ~32 km s$^{-1}$ less than the solution with the points included. In this solution, the predicted ΔRV of the two CORALIE observations is 15.9 km s$^{-1}$ less than observed.

Further short period peaks and the 800 d signal

The long period peak is at 780 d in the Generalized Lomb-Scargle periodogram (Fig 2d) with FAP = 0.095% and 790 d in the log likelihood periodogram (Fig 2e), i.e. 10 times less likely than 6.6732 d peak. No significant signals remain in the Fig 2f residual periodogram after accounting for Signals 1 and 2.

The ~800 d peak emerged only with the extension of the baseline to include the 2018 (ESO P100) observations. Considering the CHEPS data on its own or the DMPP data alone does not constrain the orbit sufficiently to make further checks on long timescales. The EMPEROR code (PAPR and JSJ; in preparation 2019) did not detect the ~800 d signal when using linear correlation terms to model the impact of the activity on the RVs. This suggests that activity effects have a small impact on DMPP-3 RVs.

Period analysis of the activity timeseries and the stellar rotation period

Phase folding on some of the prominent peaks between 10 d and 500 d confirms that they result from both window function and variability effects. In particular, the peaks in the S-index periodogram at shorter periods, are related to long timescale S-index variability apparent in the higher than average 2018 (red) points in the top panel of Fig 4. The stellar parameters in Table 1 imply the most likely rotation period of HD 42936 is $P_{rot}$ = 21.8 d (68% confidence range: 5.3 - 22.3 d). The window function confounds attempts to determine whether any of the S-index or FWHM peaks correspond to the rotation period.

Activity vs RV correlations

We also investigated the activity vs residual RV correlations after subtracting only the RV binary motion due to DMPP-3B (i.e. the solution in Supplementary Table 1). The correlation statistics for S-index, FWHM and BIS with RV residual are almost identical to the case presented in the main article, were Signal 1 is subtracted from the solution with both Signal 1 and Signal 2 (i.e. Table 2). We find respective values of $r$ = -0.03, 0.27 and 0.07 ($p$ = 0.78, 0.01 and 0.51).

**Supplementary Table 1**. Fit parameters for DMPP-3B only. See main article Table 2 for full details of tabulated parameters.

| Parameter | Value |
|---|---|
| FAP | $6.7 \times 10^{-48}$ |
| $\Delta \log L$ | 546.9 |
| $\chi_r^2$ | 1.19 |
| r.m.s. [m s$^{-1}$] | 1.38 |
| $P$ [d] | 506.837 (506.820 - 506.853) |
| $K$ [m s$^{-1}$] | 2642.6 (2611.1 - 2684.0) |
| $M_0$ [deg] | 126.169 (126.048 - 126.294) |
| $e$ | 0.5957 (0.5930 - 0.5993) |
| $\omega_0$ [deg] | 158.652 (158.477 – 158.856) |
| $\gamma_{CHEPS}$ [m s$^{-1}$] | -828.6 (-835.9 - -822.8) |
| $\gamma_{DMPP}$ [m s$^{-1}$] | -822.3 (-829.9 - -816.1) |
| $\gamma_{CORALIE}$ [m s$^{-1}$] | 3597.2 (3555.1 – 3658.1) |
| $\sigma_{CHEPS}$ [m s$^{-1}$] | 1.9 (1.9 - 3.7) |
| $\sigma_{DMPP}$ [m s$^{-1}$] | 0.8 (0.7 — 1.0) |
| $\sigma_{CORALIE}$ [m s$^{-1}$] | 0.0 (0.0 – 3.1) |
| $M_p \sin i$ [$M_J$] | 80.3 (76.9 – 83.6) |
| $a_{bin}$ [AU] | 1.222 (1.198 - 1.244) |